# Flavour violating squark and gluino decays at LHC

**Keisho Hidaka** [1]
*Department of Physics, Tokyo Gakugei University*
*Koganei, Tokyo 184-8501, Japan*
*E-mail:* `hidaka@u-gakugei.ac.jp`

**Alfred Bartl, Elena Ginina**
*University of Vienna, Faculty of Physics*
*Boltzmanngasse 5, A-1090 Vienna, Austria*

**Helmut Eberl, Walter Majerotto**
*Institut für Hochenergiephysik der ÖAW*
*A-1050 Vienna, Austria*

**Björn Herrmann**
*LAPTh, Université de Savoie, CNRS*
*9 Chemin de Bellevue, F-74941 Annecy-le-Vieux, France*

**Werner Porod**
*Institut für Theoretische Physik und Astrophysik, Universität Würzburg*
*D-97074 Würzburg, Germany*

We study the effects of squark generation mixing on squark and gluino production and decays at LHC in the Minimal Supersymmetric Standard Model (MSSM) with focus on the mixing between second and third generation squarks. Taking into account the constraints from B-physics experiments we show that various regions in parameter space exist where decays of squarks and/or gluinos into quark flavour violating (QFV) final states can have large branching ratios. Here we consider both fermionic and bosonic decays of squarks. Rates of the corresponding QFV signals, e.g. $pp \to tt\bar{c}\bar{c}$ missing-$E_T$ X, can be significant at LHC (14 TeV).

We find that the inclusion of flavour mixing effects can be important for the search of squarks and gluinos and the determination of the underlying model parameters of the MSSM at LHC.

---

[1] Speaker

## 1. Introduction

If weak scale supersymmetry (SUSY) is realized in nature, gluinos and squarks will have high production rates for masses up to O(1) TeV at LHC. The main decay modes of gluinos and squarks are usually assumed to be quark-flavour conserving (QFC). However, squark generation mixing can induce quark-flavour violating (QFV) decays of gluinos and squarks. In this article based on [1, 2] we study the effects of squark generation mixing on squark and gluino production and decays at LHC in the general Minimal Supersymmetric Standard Model (MSSM) with focus on the mixing between second and third generation squarks.

## 2. MSSM with QFV

We take the basic MSSM parameters defined at the scale Q = 1 TeV (except for $m_{A^0}$ being the pole mass) as follows: $\tan\beta$, $m_{A^0}$, $M_{1,2,3}$, $\mu$, $M^2_{Q\alpha\beta}$, $M^2_{U\alpha\beta}$, $M^2_{D\alpha\beta}$, $T_{U\alpha\beta}$ and $T_{D\alpha\beta}$ ($\alpha,\beta$ = 1,2,3 = u,c,t or d,s,b), where $\tan\beta = <H_2^0>/<H_1^0>$, $m_{A^0}$ is the CP odd Higgs boson mass, $M_{1,2,3}$ are the U(1), SU(2), SU(3) gaugino masses, $\mu$ is the higgsino mass parameter, $M^2_{Q\alpha\beta}$ is the left squark soft-SUSY-breaking mass matrix, $M^2_{U\alpha\beta}$ [$M^2_{D\alpha\beta}$] is the right up-type [down-type] squark soft-SUSY-breaking mass matrix, $T_{U\alpha\beta}$ [$T_{D\alpha\beta}$] is the trilinear coupling matrix of up-type [down-type] squarks and the Higgs bosons. The QFV parameters in our study are $M^2_{Q23}$ ($\tilde{c}_L$-$\tilde{t}_L$ mixing term), $M^2_{U23}$ ($\tilde{c}_R$-$\tilde{t}_R$ mixing term), $T_{U23}$ ($\tilde{c}_L$-$\tilde{t}_R$ mixing term) and $T_{U32}$ ($\tilde{c}_R$-$\tilde{t}_L$ mixing term). We work in the super-CKM basis of squarks. We define the following QFV parameters:

$\delta^{LL}_{23} \equiv M^2_{Q23}/\sqrt{M^2_{Q22}M^2_{Q33}}$, $\delta^{uRR}_{23} \equiv M^2_{U23}/\sqrt{M^2_{U22}M^2_{U33}}$, $\delta^{uRL}_{23} \equiv (v_2/\sqrt{2})T_{U32}/\sqrt{M^2_{U22}M^2_{Q33}}$ and $\delta^{uRL}_{32} \equiv (v_2/\sqrt{2})T_{U23}/\sqrt{M^2_{U33}M^2_{Q22}}$, where $v_2/\sqrt{2} \equiv <H_2^0>$.

## 3. Constraints on the MSSM parameters

The following constraints are taken into account in our analysis in order to respect experimental and theoretical constraints [3]: constraints from the B-physics experiments (such as $b \to s\gamma$, $\Delta M_{Bs}$ and $B_s \to \mu^+\mu^-$), LHC limits on sparticle masses [4], constraints from the LHC data on the Standard Model (SM)-like Higgs boson [5], the experimental limit on SUSY contributions to the electroweak $\rho$ parameter, and vacuum stability conditions on the trilinear couplings. Respecting the LHC limits on squark and gluino masses [4], we assume a gluino mass of about 1 TeV in our analysis.

## 4. QFV gluino 3-body decays

### 4.1 QFV Scenario

We take the following scenario as our prototype QFV scenario A (All mass parameters are in GeV.) [2]:

$(M_1, M_2, M_3) = (139, 264, 800)$, $\mu = 1000$, $\tan\beta = 10$, $m_{A^0} = 800$, $T_{U\alpha\alpha} = T_{D\alpha\alpha} = 0$ ($\alpha = 1,2,3$),

$(M^2_{Q11}, M^2_{Q22}, M^2_{Q33}) = ((3150)^2, (3100)^2, (3050)^2)$

$(M^2_{U11}, M^2_{U22}, M^2_{U33}) = ((3000)^2, (2200)^2, (2150)^2)$

$(M^2_{D11}, M^2_{D22}, M^2_{D33}) = ((3000)^2, (2990)^2, (2980)^2)$.

In this scenario the physical gluino mass (which is fairly insensitive to the QFV parameters) is $m_{\tilde{g}} = 975$ GeV for which at tree-level $\sigma(pp \to \tilde{g}\tilde{g}X) = 170$ fb at LHC (14 TeV). Note that $m_{\tilde{c}_R} \cong (M^2_{U22})^{1/2} = 2200$ GeV and $m_{\tilde{t}_R} \cong (M^2_{U33})^{1/2} = 2150$ GeV, and that all other squarks have masses of about 3 TeV. We add the $\tilde{c}_R$-$\tilde{t}_R$ mixing parameter $M^2_{U23}$ ($\sim \delta^{uRR}_{23}$) to this scenario, which can induce large mass-splitting between $\tilde{c}_R$ and $\tilde{t}_R$ resulting in a light $\tilde{u}_1$ (the lightest up-type squark). As all squarks other than $\tilde{u}_1$ are very heavy in this large $\tilde{c}_R$-$\tilde{t}_R$ mixing scenario, gluino decay is dominated by virtual exchange of $\tilde{u}_1$ which is a strong mixture of $\tilde{c}_R$ and $\tilde{t}_R$. Hence the QFV gluino 3-body decay branching ratio $B(\tilde{g} \to ct\tilde{\chi}^0_1) \equiv B(\tilde{g} \to c\bar{t}\tilde{\chi}^0_1) + B(\tilde{g} \to \bar{c}t\tilde{\chi}^0_1)$ can be very large.

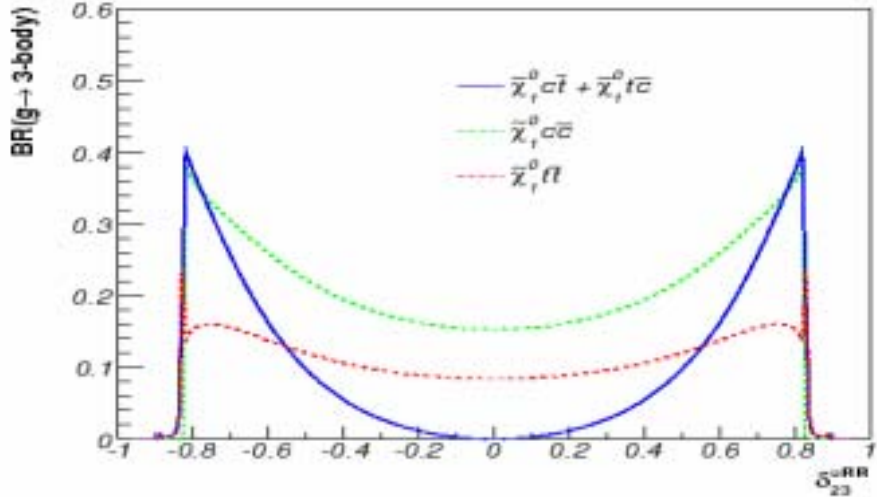

**Figure 1**: The branching ratios of the decays $\tilde{g} \to c\bar{t}\tilde{\chi}^0_1 + \bar{c}t\tilde{\chi}^0_1$, $\tilde{g} \to c\bar{c}\tilde{\chi}^0_1$ and $\tilde{g} \to t\bar{t}\tilde{\chi}^0_1$ as functions of $\delta^{uRR}_{23}$ for the scenario A with the other QFV parameters being zero.

---

[2] In this scenario we have $m_{h^0} = 121.1$ GeV. We have, however, confirmed that $m_{h^0}$ can be easily pushed up to the LHC "Higgs signal" range just by taking a sizable $T_{U33}$ without changing our final conclusion.

### 4.2 Impact of squark generation mixing on gluino 3-body decays

In Fig.1 we show the $\delta_{23}^{uRR}$ dependence of $B(\tilde{g} \to ct\tilde{\chi}_1^0)$ for the scenario A. We see that the QFV branching ratio $B(\tilde{g} \to ct\tilde{\chi}_1^0)$ can be very large (up to ~40%) for large $\delta_{23}^{uRR}$. This can lead to remarkable QFV signatures at LHC, such as
$pp \to \tilde{g}\tilde{g}X \to (t\bar{c}\tilde{\chi}_1^0)(\bar{t}c\tilde{\chi}_1^0)X \to tt\, jet\, jet\, E_T^{mis}\, X$, where $E_T^{mis}$ is missing $E_T$ and X contains beam-jets only. We find that the QFV signal rates such as
$\sigma(pp \to \tilde{g}\tilde{g}X \to (t\bar{c}\tilde{\chi}_1^0)(\bar{t}c\tilde{\chi}_1^0)X \to tt\, jet\, jet\, E_T^{mis}\, X)$ can be significant for large $\delta_{23}^{uRR}$ at LHC(14 TeV) [1].

### 5. QFV squark bosonic decays

#### 5.1 QFV Scenario

We take the following decoupling Higgs scenario as our prototype QFV scenario B (All mass parameters are in GeV.):

$(M_1, M_2, M_3) = (400, 800, 1000)$, $\mu = 2400$, $\tan\beta = 20$, $m_{A^0} = 1500$,

$(M_{Q11}^2, M_{Q22}^2, M_{Q33}^2) = ((2400)^2, (2360)^2, (1450)^2)$

$(M_{U11}^2, M_{U22}^2, M_{U33}^2) = ((2380)^2, (780)^2, (750)^2)$

$(M_{D11}^2, M_{D22}^2, M_{D33}^2) = ((2380)^2, (2340)^2, (2300)^2)$

with all of $T_{U\alpha\alpha}$ and $T_{D\alpha\alpha}$ being zero, except $T_{U33} = -2160$.

This is a decoupling Higgs scenario with a large top-trilinear-coupling $T_{U33}$ (i.e. large $\tilde{t}_L$-$\tilde{t}_R$ mixing term). In this scenario the physical masses of gluino and the lightest MSSM Higgs boson $h^0$ are $m_{\tilde{g}} = 1141$ GeV, $m_{h^0} = 125.5$ GeV, respectively. These masses are fairly insensitive to the QFV parameters. Note that in our decoupling Higgs scenario the $h^0$ is indeed SM-like and its mass $m_{h^0} = 125.5$ GeV is in the LHC "Higgs signal" range 123 GeV < $m_{h^0}$ < 129 GeV allowing for experimental and theoretical uncertainties [5, 6]. We add the $\tilde{c}_R$-$\tilde{t}_R$ mixing parameter $M_{U23}^2 (\sim \delta_{23}^{uRR})$ to this scenario, which can induce large mass-splitting between $\tilde{c}_R$ and $\tilde{t}_R$ resulting in two mass eigenstates $\tilde{u}_1$ and $\tilde{u}_2$ with a large mass difference. Here note that $m_{\tilde{c}_R} \cong (M_{U22}^2)^{1/2} = 780$ GeV and $m_{\tilde{t}_R} \cong (M_{U33}^2)^{1/2} = 750$ GeV. Hence $B(\tilde{u}_2 \to \tilde{u}_1 h^0)$ could be sizable for large $\delta_{23}^{uRR}$. Moreover, in this large $\tilde{t}_L$-$\tilde{t}_R$ and $\tilde{c}_R$-$\tilde{t}_R$ mixing scenario, we have $\tilde{u}_{1,2} \sim \tilde{c}_R + \tilde{t}_R(+\tilde{t}_L)$ and $h^0 \sim \mathrm{Re}(H_2^0)$ and hence $\tilde{u}_1$-$\tilde{u}_2$-$h^0$ coupling can be large due to the large $\tilde{t}_L$-$\tilde{t}_R$-$H_2^0$ coupling $T_{U33}$ (= -2160 GeV). This can leads to large QFV branching ratios $B(\tilde{u}_2 \to \tilde{u}_1 h^0)$ and $B(\tilde{u}_1 \to c/t\, \tilde{\chi}_1^0)$ and hence large $B(\tilde{u}_2 \to \tilde{u}_1 h^0 \to c/t\, h^0\, \tilde{\chi}_1^0)$.

### 5.2 Impact of squark generation mixing on squark bosonic decays

In Fig.2 we show the $\delta_{23}^{uRR}$ dependence of the QFV branching ratio $B(\widetilde{u}_2 \to \widetilde{u}_1 h^0)$ for the scenario B. It can be very large (up to ~ 47%) for large $\widetilde{c}_R$-$\widetilde{t}_R$ mixing parameter $\delta_{23}^{uRR}$.

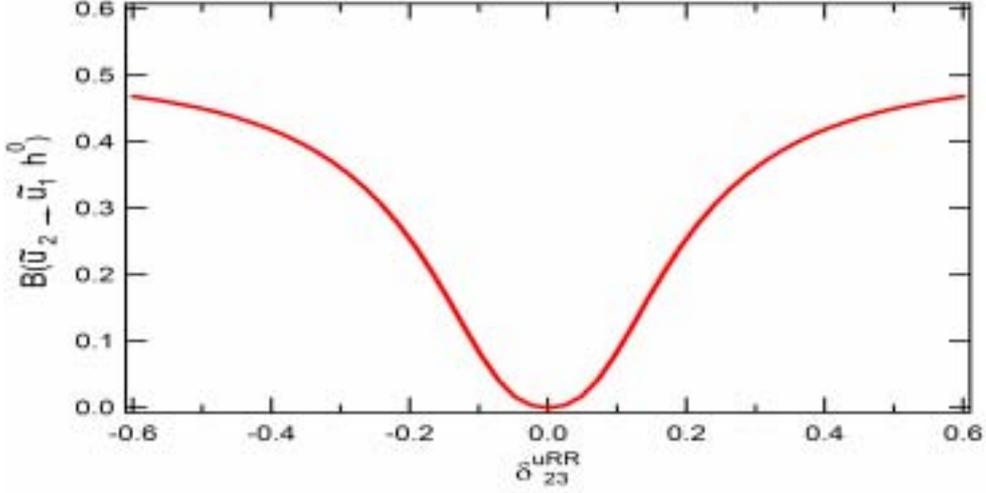

**Figure 2**: $\delta_{23}^{uRR}$ dependence of the QFV branching ratio $B(\widetilde{u}_2 \to \widetilde{u}_1 h^0)$ for the scenario B.

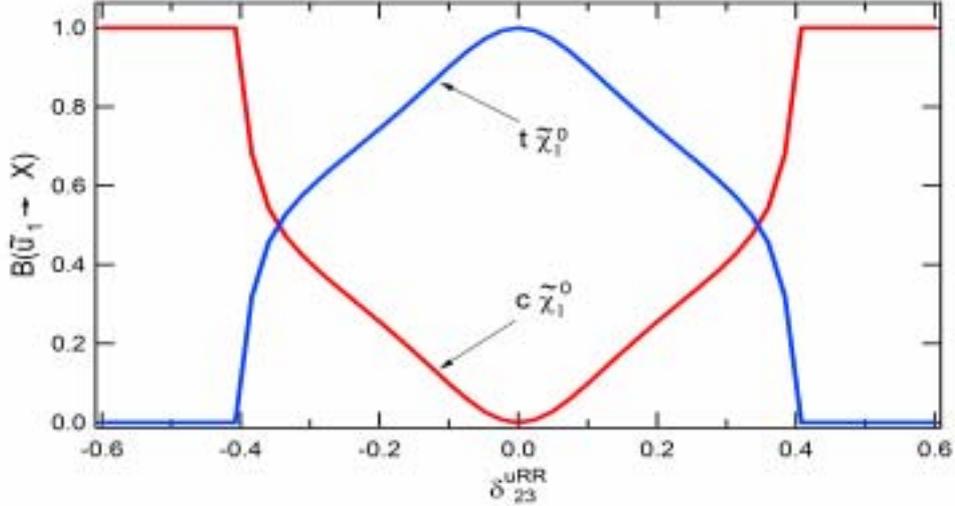

**Figure 3**: $\delta_{23}^{uRR}$ dependence of the QFV branching ratios $B(\widetilde{u}_1 \to c/t\, \widetilde{\chi}_1^0)$ for the scenario B.

In Fig.3 we show $\delta_{23}^{uRR}$ dependence of the QFV branching ratios $B(\widetilde{u}_1 \to c/t\, \widetilde{\chi}_1^0)$ for the scenario B. They can be very large simultaneously for sizable $\delta_{23}^{uRR}$, which can lead to large QFV effect.

We have also obtained a similar result for $\delta_{23}^{uRL}$ ($\widetilde{c}_R$-$\widetilde{t}_L$ mixing parameter) dependence of the QFV branching ratios $B(\widetilde{u}_2 \to \widetilde{u}_1 h^0)$ and $B(\widetilde{u}_1 \to c/t\, \widetilde{\chi}_1^0)$. In Fig.4 we show the $\delta_{23}^{uRL}$ dependence of $B(\tilde{u}_2 \to \tilde{u}_1 h^0)$ for the scenario B with $\delta_{23}^{uRR} = 0.32$ and $\delta_{23}^{LL} = 0.015$. It can be very large (up to ~50%) for large negative $\delta_{23}^{uRL}$. The increase of $B(\tilde{u}_2 \to \tilde{u}_1 h^0)$ with decrease of $\delta_{23}^{uRL}$ (~$T_{U32}$) is due to the fact that the contributions of $T_{U33}$ and $T_{U32}$ couplings to the $\tilde{u}_1$-$\tilde{u}_2$-$h^0$ coupling interfere with each other destructively (constructively) for $T_{U32} > 0$ ($T_{U32} < 0$).

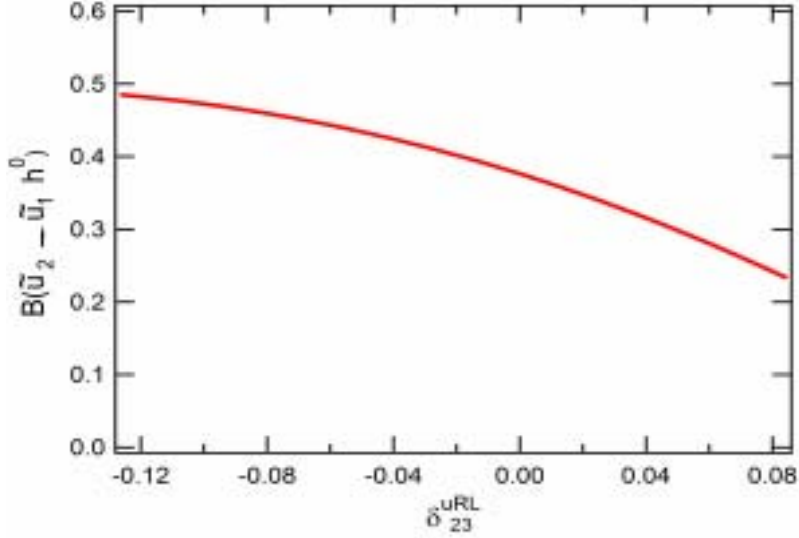

**Figure 4**: $\delta_{23}^{uRL}$ dependence of $B(\tilde{u}_2 \to \tilde{u}_1 h^0)$ for the scenario B with $\delta_{23}^{uRR} = 0.32$ and $\delta_{23}^{LL} = 0.015$.

These QFV decays can result in remarkable QFV signatures with a significant rate at LHC (14 TeV), such as $pp \to \tilde{g}\tilde{g}X \to \overline{\tilde{u}}_1 t \tilde{u}_2 \overline{c} X \to \overline{\tilde{u}}_1 t \tilde{u}_1 h^0 \overline{c} X \to \overline{c}\, \tilde{\chi}_1^0\, tc\, \tilde{\chi}_1^0\, h^0 \overline{c} X$ (=$t c \overline{c} \overline{c} h^0$ $E_T^{mis} X$). In our scenario we find that $\sigma(pp \to \tilde{g}\tilde{g}X) \sim 150$ fb at LHC(14 TeV) and that $B(\tilde{g} \to \tilde{u}_2 c/t) \equiv B(\tilde{g} \to \tilde{u}_2 \overline{c}/\overline{t}) + B(\tilde{g} \to \overline{\tilde{u}}_2 c/t)$ can be large (~ 25%), which leads to a sizable rate of $\tilde{u}_2$ production from gluino production and decays at LHC(14 TeV). Hence QFV squark signal rates can be significant at LHC(14 TeV). Indeed we have found that the QFV signal rates such as $\sigma(pp \to \tilde{g}\tilde{g}X \to t\, 3\text{jets}\, h^0\, E_T^{mis}\, X)$ can be significant for large $\delta_{23}^{uRR}$ and $\delta_{23}^{uRL}$ at LHC (14 TeV) [2].

## 6. Conclusion

Our analysis suggests the following: One should take into account the possibility of significant contributions from QFV decays in the squark and gluino search at LHC. Moreover, one should also include QFV squark parameters (i.e. squark generation mixing parameters) in the complete determination of the basic MSSM parameters at LHC.